\documentclass[aps,floats,prl,,twocolumn,floatfix]{revtex4}
\usepackage{amsmath,amssymb,graphicx,xspace,units,epsf}
\graphicspath{{figures/}}

\begin{document}
\title{Branching, Capping, and Severing in Dynamic Actin Structures}
\author{Ajay Gopinathan$^1$, Kun-Chun Lee$^2$, J. M. Schwarz$^3$ and Andrea J. Liu$^2$}
\affiliation{$^1$School of Natural Sciences, University of California, Merced,
  CA 95344 $^2$Dept.\ of Physics and Astronomy, University of Pennsylvania,
  Philadelphia, PA 19104 $^3$Dept.\ of Physics, Syracuse University, Syracuse, NY 13244}
\date{\today}
\begin{abstract}
     Branched actin networks at the leading edge of a crawling cell evolve via protein-regulated processes such
as polymerization, depolymerization, capping, branching, and
severing. A formulation of these processes is presented and
analyzed to study steady-state network morphology. In bulk, we identify several scaling regimes in severing and
branching protein concentrations and find that the coupling between severing and branching
 is optimally exploited for conditions {\it
in vivo}.  Near the leading edge, we 
find qualitative agreement with the {\it in vivo} morphology.
\end{abstract}
\maketitle
When a cell crawls, it must reorganize the cytoplasmic network of
biopolymers that controls its shape.  The shape and motility of the  
leading edge (the lamellipodium) of a
crawling cell  are determined primarily
by a dynamic network of actin filaments (F-actin) \cite{lodish}.
These filaments are living polymers made up of monomeric G-actin, and
have a definite polarity such that monomers tend to be added to the plus (or
``barbed'') end, and tend to fall off at the minus (or ``pointed'')
end.  A host of regulatory proteins concentrate plus ends at the leading
edge by controlling the morphology of the actin filament network~\cite{pollardreview}:
capping proteins bind to the plus end of filaments,
preventing further growth; severing proteins, 
such as cofilin, bind to filaments and break them in two, enhancing
depolymerization; and the protein complex Arp2/3 nucleates branches
from existing filaments, creating new plus ends at each branch tip
and protecting the bound minus ends of branches from depolymerization.
These proteins and their associated processes constitute the dendritic nucleation model~\cite{pollardreview}.  This biological model 
is supported by 
experiments that show that 
severing, capping and branching are all crucial to cell
motility~\cite{carlier,chan,bailly,aizawa,loisel}.  

In order to gain insight into how
the cell controls its structural dynamics it is imperative to
understand {\it quantitatively} how actin morphology can be controlled by the
concentrations of these regulatory proteins.  To date, 
theoretical studies of these kinetic processes 
have either omitted one or more of these critical processes
~\cite{mogilner,edelstein,carlssonPRL,carlssonBJ1,wear} or have been
restricted to describing just the overall amount of polymerization
as a function of time~\cite{carlssonBJ3}.  Other treatments of actin-based motility have focused on the interplay of force generation with motility~\cite{prost1,mogilner2},
and treat the branched actin network either as an elastic continuum or
as a collection of
uncoupled filaments.  
In this paper, we adopt a complementary approach that not only captures
all of the critical features of the dendritic nucleation model but also allows the first
theoretical investigation of the morphology of the branched network.  We confirm biological and biochemical understanding of the proteins involved by obtaining quantitative agreement with bulk {\it in vitro} experiments.   We extend the approach to explore the interplay between actin morphology and motility near the leading edge of a crawling cell.  The agreement of our theory with electron microscopy images of the leading edge of a crawling cell provides strong evidence that the dendritic nucleation model captures the key players in this important form of cell motility.

We introduce a mean-field rate equation  
formulation for
actin structures that undergo polymerization, depolymerization, capping,
branching, and severing.  We capture key morphological  
information
by retaining the entire length distributions of filaments  
with free
minus ends and of branches.  Our formulation is constructed for polar actin
assembly with branching (Arp2/3), severing (cofilin) and
capping agents, whose action is described by rate constants.
Initially, capping is taken
into account by assigning a probability for a filament to be
capped; for reversible capping, this renormalizes the
polymerization rate.  Thus,  $k_+$ is
the effective polymerization rate constant.
The growth rate at the barbed end of a filament is
$k_+ \rho_m$, where
$\rho_m$ is the free monomer concentration.  The depolymerization rate is 
denoted by $k_-$. Nucleation and dissociation of
filament seeds are described by rate constants $k_n$ and $k_{d}$  
\cite{pantaloni}
respectively.   Nucleation of branches is modeled by the rate constant $k_{arp} = k_{arp}^0 \rho_{\rm{Arp2/3}} \rho_m^2$ \cite{wear,pantaloni}, where $\rho_{\rm{Arp2/3}}$
is the concentration of Arp2/3.
Eventually, Arp2/3 dissociates and branches detach at a rate
$k_{dr}$ \cite{weaver}. Severing occurs at a rate $k_s = k_s^0 \rho_s$, where $\rho_s$ is the severing protein concentration \cite{moriyama}.
The nucleotide state of the polymerized actin controls the rates of both
branching (preferred at ATP-actin sites\cite{ichetovkin}) and severing
(preferred at ADP-actin sites\cite{pollard-borisy}).  We therefore
consider hydrolysis and
phosphate release occurring in sequence to convert newly polymerized
ATP-actin to ADP-actin at a rate $k_{pr}$ \cite{shaughnessy}.  The  
probability that a monomer at distance $L$ from the barbed end of a  
filament is an ADP-actin monomer is then given by $p(L)= 1 - \exp(-L/l_c)$ ~\cite{edelstein}, where $l_c = k_+ \rho_m/k_{pr}$.  We allow  
branching only on ATP-actin monomers and severing only on ADP-actin  
monomers in filaments.

\begin{figure}
   \centering
\hspace{-0.4cm}   \includegraphics[width=0.9\columnwidth]{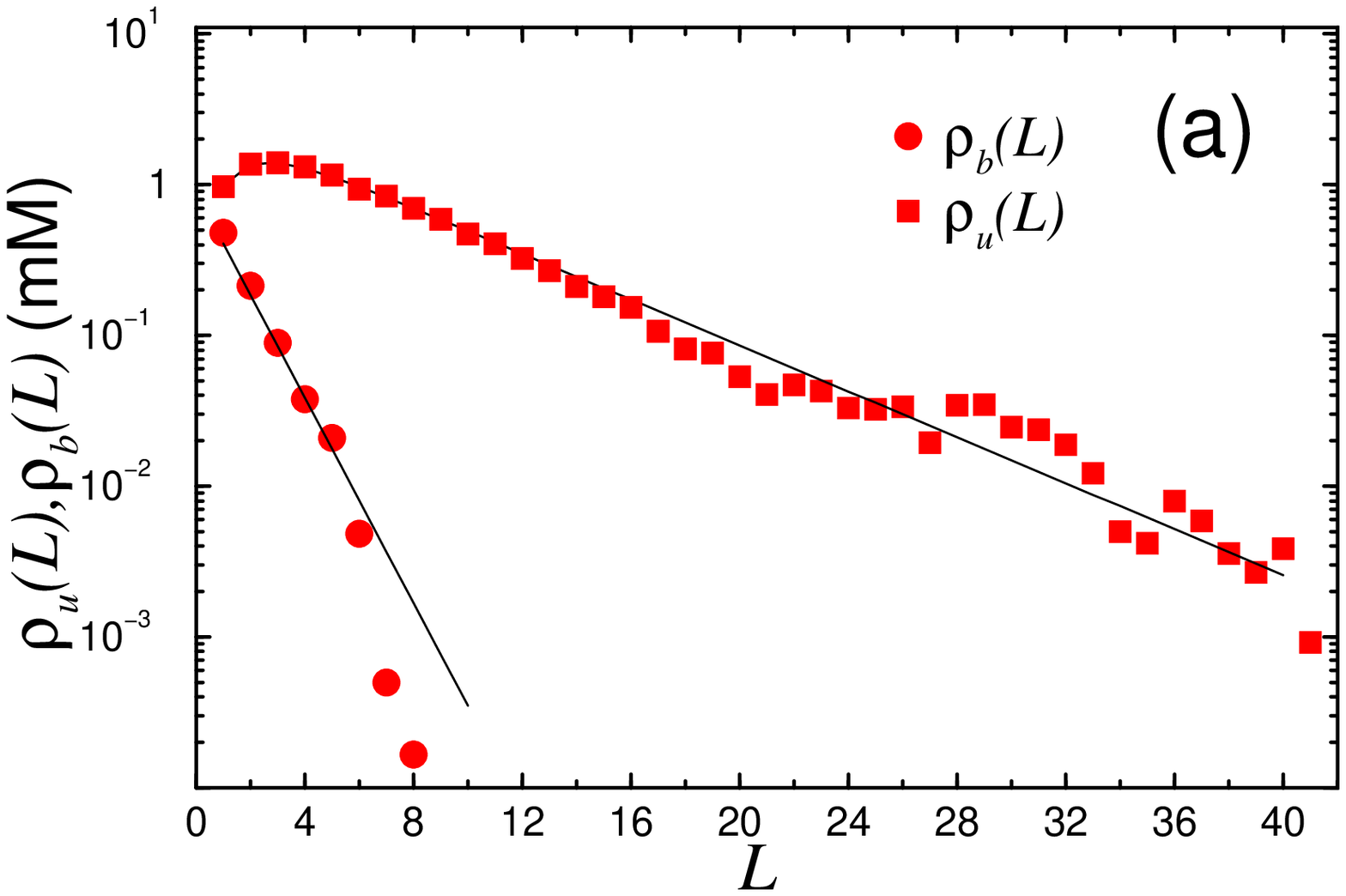}\vspace{-0.2cm}

\includegraphics[width=0.85\columnwidth]{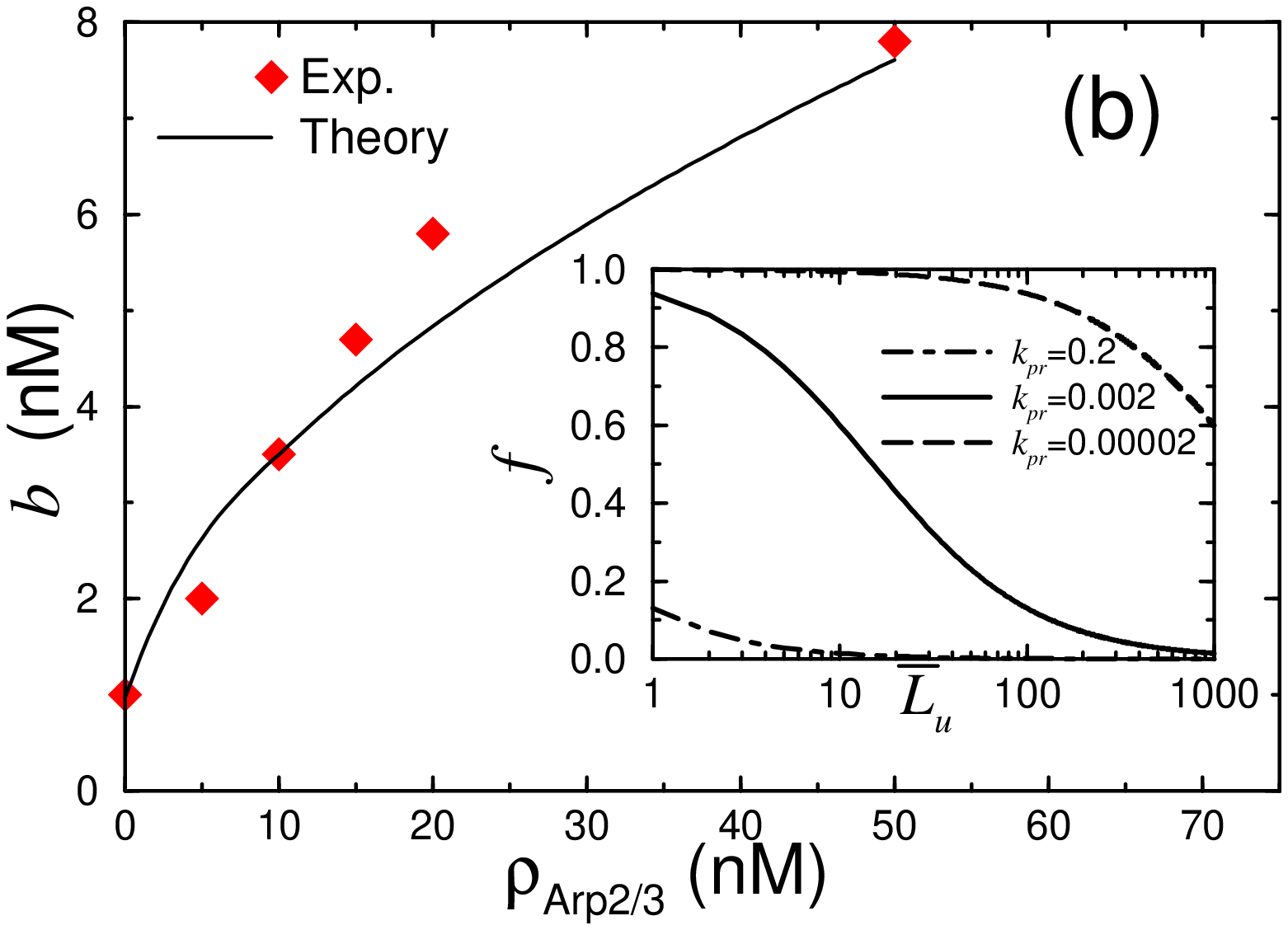}
\vspace{-0.4cm}
   \caption{(a) Length
distribution of filaments and of branches (in monomer units), as predicted by our mean  
field theory (lines) and as calculated from Brownian dynamics  
simulations (points), with $k_+=15\, \mu M^{-1}s^{-1}, k_-=k_{dr}=100\,\,s^{-1},
k_{arp}/k_+=0.015$.  (b) The barbed end concentration as a function of $\rho_{\rm{Arp2/3}}$, as predicted by mean field theory (lines) and  
measured by experiment (points)~\cite{blanchoin}.  Here, $k_n$ and $k_{arp}^0$ were 
estimated by fitting to average filament length and branching percentage data
from \cite{blanchoin} (not shown).  Inset:  The predicted steady-state value of the fraction of ATP-bound actin monomers in filaments, as a function
of $\bar{L}_u$ for various $k_{pr}$ in units of $s^{-1}$.}
\vspace{-0.5cm}
   \label{exptsim}
\end{figure}

Unless specified,  $k_+ = 8.7 \,\mu\mbox{M}^{-1}s^{-1}$ \cite{higgs2}, $k_- = 0.03\, s^{-1}$ (estimates are of the order of $10^{-2} - 10^{-1}$ \cite{kuhn} depending on conditions),
$k_{d} = 5 \times 10^{-4} \,s^{-1}$ \cite{pantaloni}, $k_n = 3 \times 10^{-6} \,\mu\mbox
{M}^{-1}s^{-1}$ (fitted; see caption of Fig.~\ref{exptsim}), $k_{arp}^0 = 4.7 \times 10^{-4} \,\mu\mbox{M}^{-3}s^{-1}$ (fitted and consistent with \cite{wear})
$k_{dr} = 0.0018 \,s^{-1}$ \cite{carlssonBJ1,wear}, $k_s^0 = 4 \times 10^{-5}
\,\mu\mbox{M}^{-1}s^{-1}$ (estimated from \cite{du,moriyama}) and 
$k_{pr} = 0.002\, s^{-1}$ \cite{melki}.
To capture the morphology we consider (1) the density of actin filaments of length $L$ (in monomer units) 
with free minus
ends ({\it i.e.} minus ends that can depolymerize), $\rho_u(L)$, (2) the
density of branches of length $L$ (which 
have bound minus ends that
cannot depolymerize), $\rho_b(L)$, and (3) the density of monomers, $\rho_m$.  In the  
mean-field bulk case where
all rate constants and densities have no spatial dependence,
we have:
\begin{widetext}
\vspace{-0.9cm}
\begin{align}
\dot{\rho_u}(L) &=  -k_+ \rho_m (\rho_u(L)-\rho_u(L-1))+k_-(\rho_u(L 
+1)-\rho_u(L)) + k_{dr} \rho_b(L) - k_s \sum_{L'=1}^{L-1} p(L') \rho_u 
(L) \notag\\ & + \sum_{L'=L+2}^{\infty} k_s p(L)\rho_b(L') +  \sum_ 
{L'=L+1}^{\infty}  k_s (p(L)+p(L'-L)) \rho_u(L') \label{rhou}\\
\dot{\rho_b}(L) &=  -k_+ \rho_m (\rho_b(L)-\rho_b(L-1)) - k_{dr} \rho_b 
(L) - k_s \sum_{L'=1}^{L-2} p(L') \rho_b(L) + \sum_{L'=L+1}^{\infty}  
k_s p(L'-L) \rho_b(L') \label{rhob}\\
\dot{\rho_u}(2) &=-k_+\rho_m\rho_u(2)+k_-\rho_u(3)-k_{d}\rho_u(2)+k_{dr}\rho_b(2)+k_n\rho_m^2+\sum_{L'=4}^ 
{\infty} k_s p(2) \rho_b(L') +\sum_{L'=3}^{\infty} k_s \tilde{p}(L') \rho_u(L')  \label{rhou_bc}\\
\dot{\rho_b}(2) &=  -k_+ \rho_m \rho_b(2) - k_{dr} \rho_b(2) + k_{arp}  
\sum_{L=2}^{\infty} (\sum_{L'=1}^{L} 1-p(L')) (\rho_u(L) +  
\rho_b(L)) + \sum_{L'=3}^{\infty} k_s p(L'-2) \rho_b(L'),
\end{align}
\label{bulkeqns} \vspace{-0.1cm}\end{widetext}
with $\tilde{p}(L')=p(2)+p(L'-2)$. A fifth equation is the conservation of total number of monomers.  These equations are similar to those in Refs.~\cite{carlssonPRL,carlssonBJ1,wear,carlssonBJ3}, though we distinguish between
unbranched and branched filaments, which is necessary to quantitatively describe morphology.

We first study the steady state in absence of severing and
nucleotide dependence.  The length
distributions are 
\begin{align}
\rho_u(L) & =  A e^{-L/\lambda_u} + B e^{-L/ \lambda_b} \label{rhou_nosev} \\
\rho_b(L) & =  C e^{-L/\lambda_b} \label{rhob_nosev} 
\end{align}
where $\lambda_u=1/\log{\frac{k_-}{k_+\rho_m}} \approx \frac{k_ 
+ \rho_m}{k_- - k_+ \rho_m}$, $\lambda_b  =  1/\log 
{\frac{k_{dr} + k_+ \rho_m}{k_+ \rho_m}} \approx \frac{k_+  
\rho_m}{k_{dr}}$.  $A$,$B$ and $C$ are length-independent and depend on the rate constants and total actin concentration.
The approximations are valid for $L>>1$, 
which is generally the case {\it in vivo}, and imply that
$k_-\approx k_+\rho_m$ and  $\lambda_b = k_+ \rho_m/k_{dr} 
\approx k_-/k_{dr}$.  

We performed Brownian dynamics simulations to determine the accuracy of Eqns.(5,6)~\cite{KC}.  We have 
used rates somewhat different from physiological ones to achieve equilibration within reasonable time.
Fig~\ref{exptsim}(a) shows that the simulations and theory yield length distributions in good agreement with each other with no adjustable parameters at these concentrations.  The predicted maximum in $\rho_{u}(L)$, which occurs because branches can fall off, is also observed in the simulations. 

We have also compared our calculations to  {\it in vitro} experiments by Blanchoin, {\it et al.} ~\cite{blanchoin}.  We neglect severing, depolymerization and nucleotide-state dependence because those experiments did not
contain severing  
proteins and filaments were stabilized by phalloidin.  
Our results for the barbed end concentration are plotted as a  
solid curve as a function of Arp2/3 concentration in Fig.~\ref
{exptsim}(b), in good agreement with the experiment. 

When severing and monomer nucleotide state are included, 
analytical progress can be made by using   
``global'' conservation principles at steady state to derive simple,
approximate expressions for morphological properties. 
In steady state, the average total number of  branches
must be conserved.  Equating the rate of production and destruction of
branches gives $k_{dr}\sum_{L=2}^{\infty} \rho_b(L)=k^0_{arp}\rho_{Arp2/3}
\rho_m^2\sum_{L=2}^{\infty}(\sum_{L'=1}^{L}(1-p(L')))(\rho_u(L)+\rho_b(L)) \approx k^0_{arp}\rho_{Arp2/3} f 
(\frac{k_{-}}{k_+})^2 \rho_m^{tot} \label{sumrhob},$ where $\rho_m^{tot}$ is the total actin concentration and  $f$ denotes the
fraction of polymerized F-actin monomers that are ATP-bound and hence
capable of supporting branching and immune to severing.

Similarly, the average number of free filaments is fixed such that  
 $k_{s} (1-f) \sum_{L=2}^{\infty} L (\rho_u(L) +
\rho_b(L))  + k_{dr} \sum_{L=2}^{\infty} \rho_b(L) = k_{d} \rho_u (2)$.
If we assume that $\rho(L)$ can be described by one
characteristic length, $\bar{L_u}$, implying $\rho_u (2)  
\approx \rho_m^{tot}/\bar{L_u}^2$, then
$\bar{L_u}= \big( \frac{k_{d}}{ k_{s}(1-f)  
+ k^0_{arp}\rho_{Arp2/3} f 
(\frac{k_{-}}{k_+})^2} \big)^{\frac{1}{2}} \label{luhalf}$. In steady  
state, conservation of ATP F-actin implies
$f \approx \big( k_{pr} \frac{\bar{L_u}}{k_-} +1 \big)^{-1}
\label{fss}$ such that
\begin{align}
\bar{L_u} \simeq \big( \frac{k_{d}(k_{pr} \bar{L_u} + k_-)}{ k_{s} 
k_{pr} \bar{L_u} + k'_{arp}(\frac{k_{-}}{k_+})^2 k_-} \big)^{\frac{1} 
{2}}, \label{lusolve}
\end{align}
where $k'_{arp}=k^0_{arp}\rho_{Arp2/3}$. 
Comparing the two terms in the denominator, we see that branching  
does not
significantly affect the average length $\bar L_u$ until the Arp2/3  
concentration, $\rho_{\rm{Arp2/3}}$,  is at least four orders of magnitude larger than the  
cofilin concentration, $\rho_s$.  {\it In vitro} motility experiments~\cite 
{carlier} typically
operate in the regime where $\rho_{\rm{Arp2/3}}<\rho_s$.  Then Eq.~\ref{lusolve} shows that for large $k_s$,
$\bar L_u$ is short
and scales as
$\bar L_u \sim (k_{d}k_-/(k_s k_{pr}))^{\frac{1}{3}}$. For lower $k_s$, (but $k_s>>k'_{arp}(\frac{k_-}{k_+})^2$),
$\bar L_u$ is larger and shows a more sensitive dependence on $k_s$:
$\bar L_u \sim (k_{d}/k_s)^{\frac{1}{2}}$.  The crossover between  
these two regimes occurs when $\bar L_u \sim 10$.   On the  
other hand, when severing is negligible and $k_{pr}$ is small such  
that side-branching dominates, then
$\bar L_u \propto 1/\sqrt{k'_{arp}}$. For small $k_s$  
and large phosphate release rates $k_{pr}$, where
end-branching dominates, $\bar L_u  \propto 1/k'_{arp}$.  These scaling predictions are experimentally testable.

The density of  
branched filaments is also constant in steady state.
As $k_{pr}$ increases, branching occurs only near the barbed 
end of filaments.  Thus, the branch density will depend not on the number density of monomers in filaments, but on the number density of {\it filaments}, which is sensitive to severing 
and is given by $\sum \rho_b  
\propto \sqrt{k_s}k'_{arp}$.  On the other hand, if there is no dependence on  nucleotide state, the branch density depends on the total F-actin concentration, which depends on $k'_{arp}$, but not sensitively on $k_s$.

These results suggest that the dependence of branch density on the  
nucleotide state of filament monomers allows the  
cell to control morphology by varying the  
concentration of severing protein \cite{carlssonBJ3}. 
For {\it in vivo} values of $k_{pr}$, the
system switches from a side-branching regime ($f=1$) to
an end-branching regime ($f=0$) as the average lengths increase from  
tens to several hundred subunits, the range typically observed {\it in vivo}.
In Fig.~\ref{exptsim}(b), we plot the predicted value of $f$ as a function of
average filament length for different values of the
phosphate release rate.  For the {\it in vivo} value of the phosphate  
release rate
and the {\it in vivo} range of filament lengths (ranging from tens to  
several hundreds of monomers),
the range of $f$ that is
covered is almost maximal. This suggests that {\it in vivo}, the branching  
morphology (controlled by $f$) is maximally  
sensitive to severing proteins.  Branching and severing have been viewed as
antagonistic, because branching promotes filament growth and
severing promotes depolymerization.  However, our results show that
all processes combine to give maximal cooperativity between branching and severing.

We now consider the coupling of morphology to motility by confining the system between
two hard walls both moving with velocity $v$.  The ``front'' wall  
models the leading edge of the cell and while the ``back'' wall  
denotes the
the back edge of the lamellipodium. Because capping is necessary to motility~\cite{loisel}, we extend Eqns. (1-4) to explicitly include a capping rate $k_c$ and uncapping rate $k_{uc}$ by defining four populations of 
filaments: capped and uncapped branched and unbranched filaments.  Now $\rho_u(L,t,z)$ denotes the density of capped or uncapped filaments of length $L$ whose  
barbed ends
are at distance $z$ from the front wall.
Experiments find that filaments are typically at an angle of $\theta=35^\circ$ with respect to the leading edge~\cite{maly}, so we assume
that all filaments are at that angle with the wall.  Because Arp2/3 is activated
at the leading edge and diffuses away from it, we model the Arp2/3
concentration profile as an exponential decay from the front 
with a decay length  $z_{arp}$~\cite{bailly2}.  Similarly, we use a severing
concentration profile that is an
inverted exponential with a rise length 
of $z_{s}$ to model the effect of nucleotide dependence on the severing efficiency.  
Finally, the amount of free monomers is determined by the  
conservation of actin, and actin that flows beyond the back wall is recycled in as free monomers; our results are for systems sufficiently large such that actin reaching the back wall is small compared to actin within the system.  We use an  
exponential profile with a decay length of $D/v_t$, consistent with the diffusion equation,
where $D$ is the diffusion constant of free  
monomers and $v_t$ is the transport (bias) velocity
of free monomers towards the surface.  
We choose $v_t=250\,v$, where  
$v$ is the speed of the moving surface, consistent with 
results of Zicha {\it et al.}~\cite{zicha}. 
Not all values of $v_t$ yield reasonable morphologies.  If, for example, $v_t=500\, v$, the system tends toward a state with no filaments. 

\begin{figure}
\centering
\hspace{-0.4cm}\includegraphics[width=0.9\columnwidth]{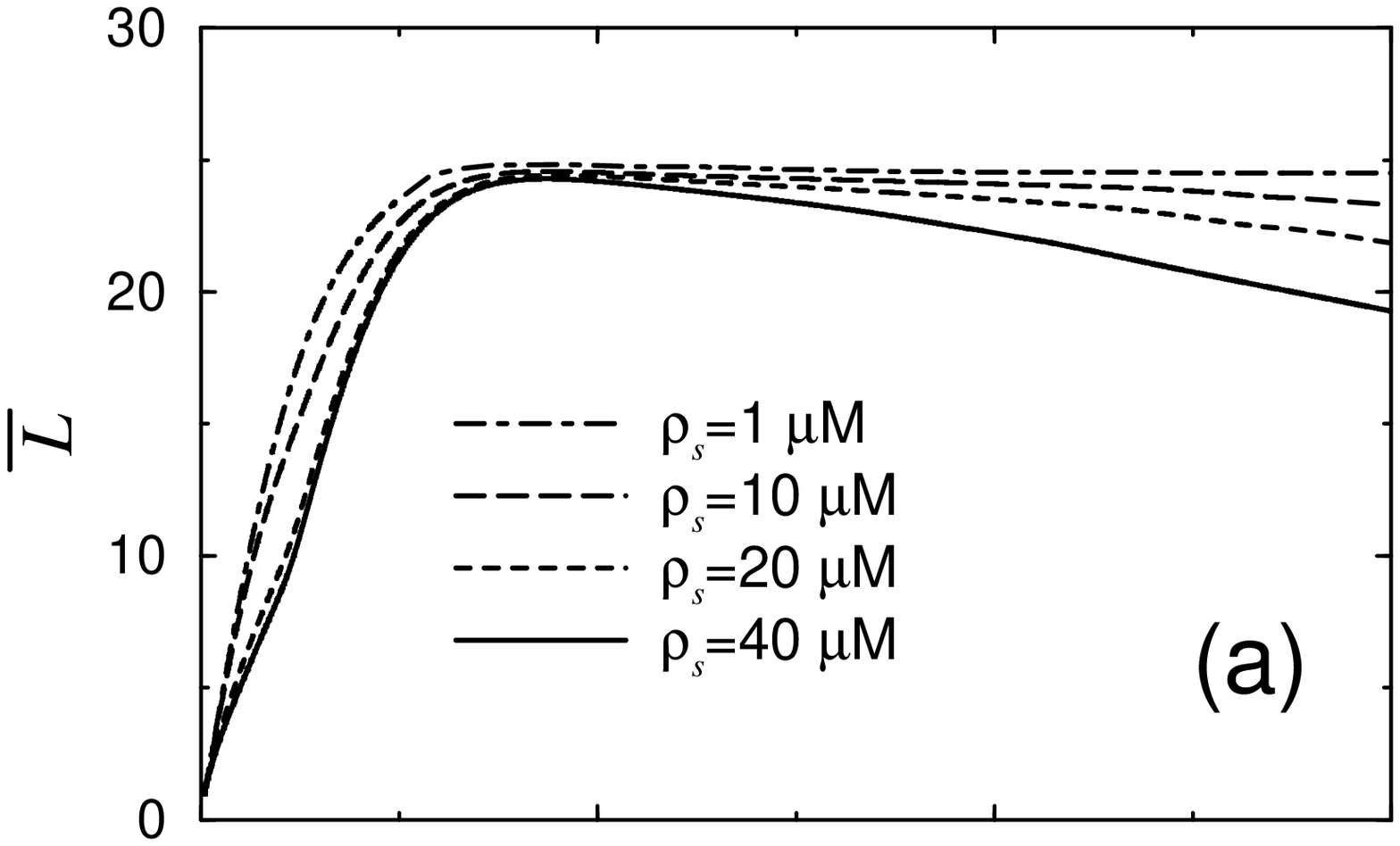}
\includegraphics[width=0.9\columnwidth]{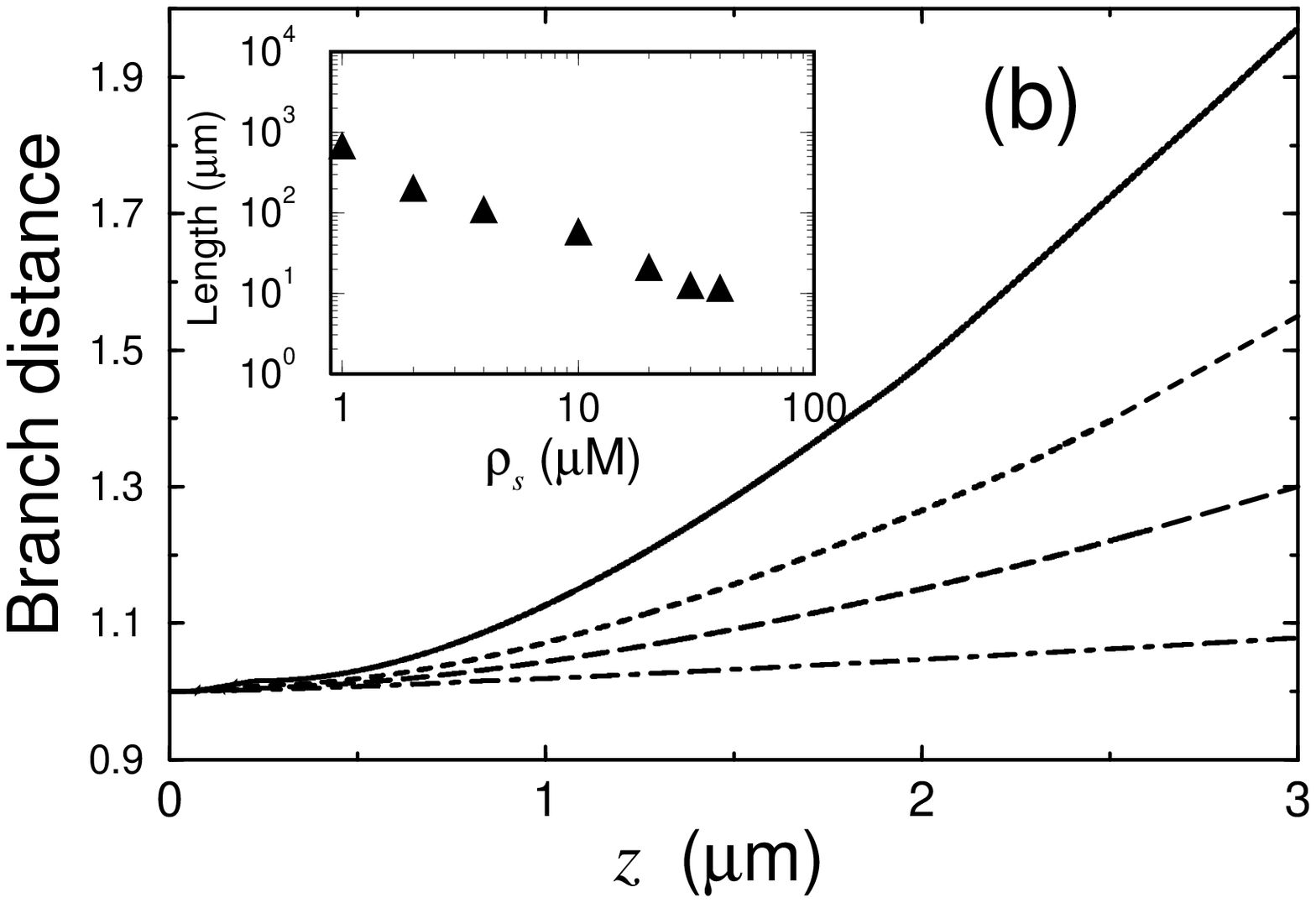}
\vspace{-0.4cm}
   \caption{(a) $\bar L$  {\it vs.} distance from leading edge, $z$, for different cofilin 
concentrations, $\rho_s$. Here, $v=0.1 \,\mu$m/min, $v_t=25\,\mu$m/$s$, $D=5\,\mu$m$^2/s$, $z_{arp}=0.1 \,\mu$m, $z_{s}=2\,\mu$m,$ \,k_c=1 \,s^{-1}$, $k_{uc}=0.01 \,s^{-1}$ with 1 $\,\mu$M Arp2/3 and 100 $\,\mu$M actin (at the upper end of the physiological range).  (b) Distance between branches {\it vs.} $z$ for the same $\rho_s$ as in (a).  Inset: Lamellipodium length {\it vs.} 
$\rho_s$.  }
   \label{surfacefig}
\vspace{-0.5cm}
\end{figure}

In steady-state, we find that branching, severing, 
and capping are all needed to obtain a morphology that is consistent with 
experiments. Branching in the front is needed to help the system ``keep up'' with the wall 
by nucleating new filaments.  Capping towards the front channels new filament growth into shorter branches and thus increases filament density.  Finally, severing is needed to replenish the free monomer supply. 
Fig. ~\ref{surfacefig} shows the steady state morphology for the system with a  
moving surface. In Fig.~\ref{surfacefig}(a), we plot the
average length of filaments (branched and unbranched) as a function of distance $z$ from the  
front wall.  The average length, $\bar L$, is
short near the surface, and increases to approximately 100 nm at a  
distance of roughly $0.5\,{\mu}$m from the surface.  As the severing protein concentration $\rho_s$ increases, $\bar L$  
decreases with increasing $z$, eventually  
reaching zero when all filaments are depolymerized 
away.  The distance at which $\bar L$ reaches zero provides an upper bound on
the lamellipodium length, {\it i~.e~.} the distance the lamellipodium extends into the cell from the leading edge.  This length is  
plotted in the inset to Fig.~\ref{surfacefig}(b) as a function of $\rho_s$.  For tens of micromolar concentrations, the  
lamellipodium length is tens of microns, consistent with experimental  
observations~\cite{svitkina1}.   The predicted length increases with actin concentration.

Fig.~\ref{surfacefig}(b) shows that the 
average distance between branches monotonically increases with $z$.  
The higher $\rho_s$, the  
greater the increase in the branch distance with $z$.
In short, Fig.~\ref{surfacefig} implies that the morphology  
consists of short branched filaments within the first micron of the  
leading edge, with longer, less branchy filaments further  
away.   These observations are consistent with electron microscopy  
images of the lamellipodium in crawling keratocytes~\cite{svitkina1},
suggesting that the dendritic nucleation model captures the minimal set of proteins involved. 
A thorough, quantitative experimental test 
of our results near a moving surface should provide a stringent 
check on the validity  
of the dendritic
nucleation model for actin assembly near the leading edge of a  
crawling cell. 

We thank T. Svitkina for instructive discussions and are
grateful for support from NSF-CHE-0613331 and NSF-DMR-0520020. 


\vspace{-0.5cm}
\begin{thebibliography}{99}
\vspace{-0.5cm}
\bibitem{lodish}  H. Lodish, {\it et al.}, {\it Molecular Cell  
Biology}, 3rd ed., (W. H. Freeman, New
York, NY, 1995).
\bibitem{pollardreview}T. D. Pollard, {\it et al.}, Annu. Rev. Biophys. Biomol. Struct. {\bf 29}, 545 (2000).
\bibitem{carlier} D. Pantaloni, {\it et al.}, Science  
{\bf 292}, 1502 (2001).
\bibitem{chan} A. Y. Chan,{\it et al.}, J. Cell. Biol. {\bf 148}, 531  
(2000).
\bibitem{bailly} M. Bailly, {\it et al.}, Curr. Biol. {\bf 11}, 620  
(2001).
\bibitem{aizawa} H. Aizawa, {\it et al.}, J. Cell. Biol.  
{\bf 132}, 335 (1996).
\bibitem{loisel} T. Loisel, {\it et al.}, Nature {\bf 401}, 613 (1999).  
\bibitem{mogilner} A. Mogilner and L. Edelstein-Keshet, Biophys. J.  
{\bf 83}, 1237 (2002).
\bibitem{edelstein} L. Edelstein-Keshet and G. Bard Ermentrout, J.  
Math. Biol. {\bf 43}, 325 (2001).
\bibitem{carlssonPRL} A. E. Carlsson, Phys. Rev. Lett., {\bf 92},  
238102 (2004).
\bibitem{carlssonBJ1} A. E. Carlsson, Biophys. J. {\bf  
89}, 130 (2005).
\bibitem{wear} A.E. Carlsson, {\it et al.}, Biophys. J.  
{\bf 86}, 1074 (2004).
\bibitem{carlssonBJ3} A. E. Carlsson, Biophys. J. {\bf 90}, 413 (2006).
\bibitem{prost1} K. Kruse {\it et al.}, Phys. Biol. {\bf 3}, 130 (2006).

\bibitem{mogilner2} A. Mogilner and G. Oster, Biophys J. {\bf 84}, 1591 (2003).

\bibitem{pantaloni} D. Pantaloni, {\it et al.}, Nature Cell Biol. {\bf 2}, 385 (2000).




\bibitem{weaver} A.M. Weaver {\it et al.}, Curr. Biol. {\bf 11}, 370  
(2001).

\bibitem{moriyama} K. Moriyama and I. Yahara, EMBO J. {\bf 18},  
6752 (1999).
\bibitem{ichetovkin} I. Ichetovkin, {\it et al.}, Curr. 
Biol. {\bf 12}, 79 (2002).
\bibitem{pollard-borisy} T.D. Pollard and G.G. Borisy, Cell {\bf  
112}, 453 (2003).
\bibitem{shaughnessy} D. Vavylonis, {\it et al.}, 
PNAS {\bf 102}, 8543 (2005).
\bibitem{higgs2} H.N. Higgs, {\it et al.},  Biochemistry {\bf 38}, 15212(1999) 
\bibitem{kuhn} J.R. Kuhn and T.D. Pollard, Biophys. J.  
{\bf 88}, 1387 (2005).
\bibitem{du} J. Du and C. Frieden, Biochemistry {\bf 37}, 13276  
(1998).
\bibitem{melki} R. Melki, {\it et al.}, Biochemistry {\bf 35},  12038 (1996) 


\bibitem{KC} K.-C. Lee and Andrea J. Liu (unpublished).
\bibitem{blanchoin} L. Blanchoin, {\it et al.}, Nature {\bf 404}, 1007  
(2000).
\bibitem{maly} I. Maly and G. Borisy, PNAS {\bf 98}, 11324 (2001).
\bibitem{bailly2} M. Bailly, {\it et al.} J. Cell. Biol. {\bf 145}, 221 (1999). 
\bibitem{zicha} D. Zicha, {\it et al.}, Science {\bf 300}, 142 (2003).
\bibitem{svitkina1} T. Svitkina, {\it et al.}, J. Cell  
Biol. {\bf 145}, 1009 (1999).
\end{thebibliography}
\end{document}